\documentclass[12pt]{article}

\usepackage{mathtools}
\usepackage{booktabs}
\usepackage[english]{babel} 
\usepackage[protrusion=true,expansion=true]{microtype} 
\usepackage{amsmath,amsfonts,amsthm}
\usepackage{ amssymb }
\usepackage{enumitem}
\usepackage{graphicx}
\usepackage{array}
\usepackage[toc,page]{appendix}
\usepackage{xcolor}

\usepackage{tikz}
\usetikzlibrary{arrows.meta, positioning}
\usetikzlibrary{matrix, positioning}
\usepackage{booktabs} 
\usepackage{natbib}
\usepackage{setspace}
\usepackage{microtype} 
\usepackage{tabularx}
\usepackage{subcaption}
\usepackage[font = small,labelfont=bf,textfont=it]{caption} 
\usepackage{footnote}
\usepackage{algorithm}
\usepackage{algpseudocode}
\usepackage{multirow}
\usepackage{hyperref}

\makeatletter
\newcommand{\distas}[1]{\mathbin{\overset{#1}{\kern\z@\sim}}}%
\newsavebox{\mybox}\newsavebox{\mysim}

\theoremstyle{definition}

\newcommand{\distras}[1]{%
  \savebox{\mybox}{\hbox{\kern3pt$\scriptstyle#1$\kern3pt}}%
  \savebox{\mysim}{\hbox{$\sim$}}%
  \mathbin{\overset{#1}{\kern\z@\resizebox{\wd\mybox}{\ht\mysim}{$\sim$}}}%
}
\newcolumntype{C}[1]{>{\centering\let\newline\\\arraybackslash\hspace{0pt}}m{#1}}

\newcommand{\blind}{1}

\begin{document}

\def\spacingset#1{\renewcommand{\baselinestretch}%
{#1}\small\normalsize} \spacingset{1.3}


\if1\blind
{
 \centering{\bf\Large Fast and accurate emulation of complex dynamic simulators}\\
  \vspace{0.3in}
  \centering{Junoh Heo\footnote{Corresponding author. 
  }\footnote{These authors gratefully acknowledge funding from NSF DMS 2338018.}\vspace{0in}\\
        Michigan State University\\
        }
    \date{\vspace{-7ex}}
} \fi

\if0\blind
{
  \bigskip
  \bigskip
  \bigskip
    \begin{center}
    {\LARGE\bf }
\end{center}
  \medskip
} \fi

\bigskip
\begin{abstract}
Dynamic simulators are computational models governed by differential equations that evolve over time. They are essential for scientific and engineering applications but remain challenging to emulate because of the unpredictable behavior of complex systems. To address this challenge, this paper introduces a fast and accurate Gaussian Process (GP)-based emulation method for complex dynamic simulators. By integrating linked GPs into the one-step-ahead emulation framework, the proposed algorithm provides exact and tractable computation of the posterior mean and variance, solving a problem previously considered computationally intractable and eliminating the need for expensive Monte Carlo approximations. This approach substantially reduces computation time while maintaining or improving predictive accuracy. Furthermore, the method naturally extends to systems with forcing inputs by incorporating them as additional variables within the GP framework. Numerical experiments on multiple dynamic systems demonstrate the efficiency and computational advantages of the proposed approach. An \textsf{R} package, \textsf{dynemu}, which implements the one-step-ahead emulation approach, is available on \textsf{CRAN}.
\end{abstract}

\noindent%
{\it Keywords}: Surrogate model, Linked Gaussian processes, Complex system, Uncertainty propagation
\vfill

\newpage
\spacingset{1.45} 

\section{Introduction}\label{sec:intro}

Computer models, often referred to as computer simulators, are indispensable tools for simulating complex systems and processes to gain insights into physical or scientific phenomena. Many of these models evolve over time and are governed by intricate systems of differential equations, collectively known as dynamic simulators \citep{scheinerman2012invitation}. Dynamic simulators are used to describe a wide range of phenomena across diverse fields, including meteorology \citep{millan2009meteorological}, hydrology \citep{li2016seasonal}, physics \citep{bahamonde2018dynamical}, engineering \citep{bongard2007automated}, and biology \citep{hwang2025bayesian}, among others. These models are particularly crucial for studying problems where analytical solutions to the governing equations are either impractical or infeasible.


Despite their utility, dynamic simulators often come with significant computational demands, which can limit their practicality in certain applications. These models typically require solving systems of ordinary differential equations (ODEs), which can be computationally intensive, especially when simulations must be repeated for tasks such as optimization, sensitivity analysis, or uncertainty quantification. To overcome these limitations, emulation offers an efficient alternative by constructing surrogate models that approximate the behavior of the simulator. Surrogate models emulate the simulator’s outputs at a fraction of the computational cost, enabling researchers to conduct extensive analyses without the prohibitive expense of repeated direct simulations. This capability enhances the utility of dynamic simulators, facilitating faster and more scalable solutions to complex problems across scientific and engineering domains.

Gaussian processes (GPs) \citep{rasmussen2006gaussian} are a widely used choice for constructing emulators due to their flexibility and strong theoretical foundation. They are well-suited for approximating complex, nonlinear relationships commonly encountered in real-world problems, although their performance can be sensitive to the choice of kernel and modeling assumptions. A key advantage of GPs lies in their ability not only to provide mean predictions but also to quantify uncertainty through analytical posterior expressions. In deterministic settings, GPs interpolate training data, a property particularly important for dynamic systems governed by deterministic equations, as it ensures consistency between the emulator and the original simulator outputs.

Several GP emulators for dynamic simulators have been developed, ranging from classic methods to recent advancements \citep{girard2002gaussian, bhattacharya2007simulation, conti2009gaussian, mohammadi2024emulating}. For a detailed overview of dynamic simulator emulation, we refer readers to \cite{mohammadi2024emulating}. Notably, \cite{bhattacharya2007simulation} proposed a one-step-ahead emulation approach, which assumes that the model output at time $t$ depends only on the output at the previous time point $t-1$, a property commonly referred to as the Markov property. Building on this, \cite{mohammadi2019emulating} enhanced the approach by incorporating input uncertainty and emulating the numerical flow map. Despite these advancements, many emulators still require intensive computations, such as Monte Carlo (MC) approximations \citep{conti2009gaussian, mohammadi2019emulating} or grid-based methods \citep{bhattacharya2007simulation}. 

To the best of our knowledge, the posterior distribution of GPs with uncertain inputs is known to be intractable, necessitating the use of MC methods \citep{girard2002gaussian, mohammadi2019emulating} in the context of dynamic simulators. To address this limitation, \cite{mohammadi2024emulating} recently employed random Fourier features to approximate the kernel. However, maintaining the quality of the approximated kernel requires drawing hundreds of random features. Furthermore, generating predictions involves drawing hundreds of additional sample paths from the emulated flow map. These computational demands may limit the practical advantages of emulators, highlighting the need for more efficient approaches.

In this article, we propose a fast and accurate emulation method for dynamic simulators by treating the emulator as self-coupled and adopting a one-step-ahead approach, thereby addressing the limitations of MC approximations. The proposed method builds on the framework of \cite{mohammadi2019emulating} but replaces the MC-based estimation of the posterior with exact closed-form expressions for the predictive mean and variance. These analytically tractable expressions are enabled by linked GPs \citep{kyzyurova2018coupling, ming2021linked}, originally developed for coupled computer models. In the context of dynamic emulation, recursive prediction introduces uncertainty at each step, making the inputs to the emulator uncertain inputs (i.e., stochastic inputs represented as random variables) rather than fixed values. The linked GP formulation allows this uncertainty to be propagated analytically over time, eliminating the need for repeated sampling and significantly reducing computational cost. By leveraging this framework, our method achieves computational efficiency without compromising accuracy, as most of the computational burdens in existing methods arise from the approximation step.

The remainder of this article is organized as follows. Section \ref{sec:review} provides a review of GP emulators, linked GPs, and the one-step-ahead approach. The proposed fast and accurate emulation method is introduced in Section \ref{sec:emulation}. Section \ref{sec:studies}, presents several numerical studies to demonstrate the competitiveness of the proposed method. Finally, Section \ref{sec:conclusion} concludes the paper.

\section{Review}\label{sec:review}

\subsection{Gaussian processes}

Let $f: \mathbb{R}^d \rightarrow \mathbb{R}$ denote a deterministic black-box function representing the output of a computer model. The input design matrix is given by $X = (\mathbf{x}_1, \ldots, \mathbf{x}_n)^\top \in \mathbb{R}^{n \times d}$, where each $\mathbf{x}_i \in \mathbb{R}^d$ is a design point. The corresponding outputs are collected in the vector $\mathbf{y}=(y_1,\ldots,y_n)^\top$ with each response defined by $y_i=f(\mathbf{x}_i)$. Typically, a GP prior assumes that the simulation outputs $\mathbf{y}$ follow a multivariate normal distribution:
\begin{align*}
    \mathbf{y} \sim \mathcal{N}_n \left( \alpha(\cdot), \tau^2 K(\cdot, \cdot) \right),
\end{align*}
where $\alpha(\cdot)$ is the mean function, $\tau^2$ is a scale parameter, and $K(\cdot, \cdot')$ is a positive-definite kernel function. Popular choices for the mean function $\alpha(\mathbf{x})$ include linear $(1, \mathbf{x}^\top) \beta$, constant $\alpha$, or zero, where $\beta$ is a vector of $d+1$ regression coefficients. In this paper, we adopt the linear mean function $\alpha(\mathbf{x})=(1, \mathbf{x}^\top) \beta$. For the kernel $K$, the squared exponential kernel or Mat\'ern kernel \citep{stein1999interpolation} is commonly adopted. In this paper, we focus on the anisotropic squared exponential kernel $K(\mathbf{x},\mathbf{x}')=\exp{\left(-\sum^d_{k=1} \frac{\left(x_k-x_k'\right)^2}{\theta_k}\right)}$, where $\mathbf{x}=(x_1,\ldots,x_d)^\top$ and $\mathbf{x}'=(x_1',\ldots,x_d')^\top$ are input vectors of size $d \times 1$, and $(\theta_1, \ldots, \theta_d)$ are the lengthscale hyperparameters representing the rate of correlation decay in each input dimension.

The posterior prediction of GPs at a new input point $\mathbf{x}$, given the input design $X$ and corresponding outputs $\mathbf{y}$, is given as:
\begin{align*}
    \mu(\mathbf{x}) &= \alpha(\mathbf{x}) + \mathbf{k}^\top(\mathbf{x}) \mathbf{K}^{-1} \left(\mathbf{y}-\alpha(X)\right), \quad \text{and}\\
    \sigma^2(\mathbf{x}) &= \tau^2 \left(1 - \mathbf{k}^\top(\mathbf{x}) \mathbf{K}^{-1} \mathbf{k}(\mathbf{x}) \right),
\end{align*}
where $\mathbf{k}(\mathbf{x})_i = K(X_i, \mathbf{x})$ and $\mathbf{K}_{ij}=K(\mathbf{x}_i,\mathbf{x}_j)$. For more details about GPs in the context of computer experiments, we refer to \cite{santner2018design} and \cite{gramacy2020surrogates}.

The model parameters $\left\{ \beta, \tau^2, \theta_1, \ldots, \theta_d \right\}$ are estimated by maximizing the log-likelihood:
\begin{align*}
    -\frac{1}{2} \left( n \log 2\pi + n\log \tau^2 + \log |\mathbf{K}| + \frac{1}{\tau^2} \left(\mathbf{y}-\alpha(X)\right)^\top \mathbf{K}^{-1} \left(\mathbf{y}-\alpha(X)\right)  \right).
\end{align*}
The profile log-likelihood is obtained by substituting the analytical expressions of $\hat{\beta}$ and $\hat{\tau}^2$: 
\begin{align*}
    \hat{\beta} = \left(X_n^\top \mathbf{K}^{-1} X_n\right)^{-1} X_n^\top \mathbf{K}^{-1} \mathbf{y} \quad \text{and} \quad \hat{\tau}^2 = \frac{\left(\mathbf{y}-(1, X_n^\top)\hat{\beta}\right)^\top \mathbf{K}^{-1} \left(\mathbf{y} - (1, X_n)\hat{\beta}\right)}{n}.
\end{align*}
All hyperparameters in this paper are estimated by maximum likelihood using the \textsf{L-BFGS-B} algorithm \citep{byrd1995limited, zhu1997algorithm}.

\subsection{Linked GPs}

Linked GP emulators \citep{kyzyurova2018coupling, ming2021linked} are designed to emulate two or more coupled systems of computer models. Consider $n$ computer models at the first layer, $f_l$ for $l=1,\ldots,n$, which are connected to a second-layer model $g$. In this framework, the outputs from $f_1, \ldots, f_{n}$ serve as part of the input to $g$. Assuming $f_1, \ldots, f_{n}$ are independent, the posterior predictive distribution at a new input $\mathbf{x}=(\mathbf{x}_1, \ldots, \mathbf{x}_n, \mathbf{x}_g)$ where $\mathbf{x}_l$ is the input to $f_l$ and $\mathbf{x}_g$ is the unlinked input to $g$, is expressed as:
\begin{align}\label{eq:posterior}
    p&(g \circ (f_1, \ldots, f_{n}) (\mathbf{x}))  =\int p(g \left(\mathbf{x}_g, f_1(\mathbf{x}_1), \ldots, f_{n}(\mathbf{x}_n) \right) )  \prod^{n}_{l=1} p(f_l(\mathbf{x}_l)) df_1(\mathbf{x}_f) \ldots df_{n}(\mathbf{x}_f).
\end{align}
This posterior is known to be analytically intractable \citep{girard2002gaussian}. Numerical techniques, such as Markov Chain Monte Carlo, can approximate the posterior but are computationally demanding. To address this limitation, subsequent works replaced $f_1, \ldots, f_{n}$ and $g$ with their respective GP emulators $\hat{f}_1, \ldots, \hat{f}_{n}$ and $\hat{g}$ and applied Gaussian approximations. Using this approach, \cite{kyzyurova2018coupling} derived the closed-form expressions for the posterior mean and variance under the squared exponential kernel, while \cite{ming2021linked} extended these results to Mat\'ern kernels. 

Linked GP emulators minimize the Kullback-Leibler divergence between the emulator and a Gaussian density \citep{ming2021linked}. Moreover, they demonstrate superior performance compared to the composite emulator, which disregards the coupled relationships and considers only the inputs of the first layer and the outputs of the final layer. This concept of linked GPs can be naturally extended to systems with multiple layers involving iterative or parallel structures \citep{ming2021linked, heo2023active}, high-dimensional output \citep{dolski2024gaussian}, as well as deep GPs \citep{ming2023deep}.

\subsection{One-step-ahead emulations}\label{sec:onestep}

Consider a dynamic computer simulator $f$ governed by a set of $d$ ODEs, where the state at time $t_s$ is denoted by $\mathbf{x}(t_s) = (x_1(t_s),\ldots,x_d(t_s))^\top \in \mathbb{R}^d$, for $s=0,\ldots,T$. The simulator may also include $p$-dimensional forcing inputs $\mathbf{w}(t_s)=(w_1(t_s),\ldots,w_{p}(t_s))^\top$, which are typically assumed to be known. The simulator maps the current state and forcing inputs to the next state, 
\begin{align*}
f(\mathbf{x}(t_s),\mathbf{w}(t_s))
=
\begin{bmatrix}
f_1(\mathbf{x}(t_s),\mathbf{w}(t_s))\\
\vdots\\
f_d(\mathbf{x}(t_s),\mathbf{w}(t_s))
\end{bmatrix}
=
\begin{bmatrix}
x_1(t_{s+1})\\
\vdots\\
x_d(t_{s+1})
\end{bmatrix}
= \mathbf{x}(t_{s+1}),
\end{align*}
thereby generating $d$ output trajectories over $T$ time steps. 

The one-step-ahead approach emulates the system's flow map over short time intervals, assuming the Markov property: the next state $\mathbf{x}(t_{s+1})$ depends only on the current state $\mathbf{x}(t_s)$ and the forcing input $\mathbf{w}(t_s)$. Once a set of emulators $\{\hat{f}_m\}_{m=1}^d$ is constructed using training data $\{(\mathbf{x}^i(t_0),\mathbf{w}^i(t_0)), \mathbf{x}^i(t_1)\}_{i=1}^n$, future states $\mathbf{x}(t_{s+1})$ can be predicted recursively by leveraging the Markov assumption \citep{mohammadi2019emulating, mohammadi2024emulating}. Importantly, this one-step-ahead approach requires only pairs of states one step apart rather than full trajectories, thereby substantially reducing the need to solve ODEs across all time steps. For notational simplicity, we write $\mathbf{x}(t_{s})$ for predicted states for $s \geq 1$. Under this convention, the recursive prediction is given by:
\begin{align}\label{eq:markov}
    x_m(t_{s+1})=\hat{f}_m(\mathbf{x}(t_s),\mathbf{w}(t_s)),
\end{align} 
for $m=1,\ldots,d$, and $s=0,\ldots,T-1$. This approach is particularly effective for periodic systems, as it can accurately predict the next state based on the current state within the input range, making it well-suited for capturing repetitive dynamics. Furthermore, the one-step-ahead approach provides an efficient, reliable, and computationally inexpensive alternative to other emulation methods for dynamic simulators \citep{conti2009gaussian, stolfi2021emulating}.

\cite{mohammadi2019emulating} extends this framework by incorporating input uncertainty at each time step by adopting MC methods, and approximating the posterior in \eqref{eq:posterior} via the laws of total expectation and total variance:
\begin{align}
    \mathbb{E}\left[ x_m(t_{s+1}) \right]&= \mathbb{E}\left[ \hat{f}_m(\mathbf{x}(t_{s}),\mathbf{w}(t_s)) \right]= \mathbb{E}_{\mathbf{x}(t_{s})}\left[\mathbb{E}\left[ \hat{f}_m(\mathbf{x}(t_{s}),\mathbf{w}(t_s)) | \mathbf{x}(t_{s}),\mathbf{w}(t_s) \right]\right], \label{eq:totalexp}\\ 
    \mathbb{V}\left[ x_m(t_{s+1}) \right]&=\mathbb{V}\left[ \hat{f}_m(\mathbf{x}(t_{s}),\mathbf{w}(t_s)) \right] \nonumber \label{eq:totalvar} \\
    &=\mathbb{E}_{\mathbf{x}(t_{s})}\left[\mathbb{V}\left[ \hat{f}_m(\mathbf{x}(t_{s}),\mathbf{w}(t_s)) | \mathbf{x}(t_{s}),\mathbf{w}(t_s) \right]\right] +\mathbb{V}_{\mathbf{x}(t_{s})}\left[\mathbb{E}\left[ \hat{f}_m(\mathbf{x}(t_{s}),\mathbf{w}(t_s)) | \mathbf{x}(t_{s}),\mathbf{w}(t_s) \right]\right].
\end{align}
In practice, equations \eqref{eq:totalexp} and \eqref{eq:totalvar} are approximated by drawing $n_{\mathrm{MC}}$ random samples of the uncertain inputs at every prediction step and evaluating the GP emulator at each sample. This procedure is repeated sequentially across all $T$ prediction steps, causing the total computational cost to increase with $n_{\mathrm{MC}}$ and $T$. While this framework effectively propagates input uncertainties through the emulator and has been extended to account for correlated emulators, \cite{mohammadi2019emulating} considers only systems without forcing inputs, limiting its flexibility for simulators with external covariates.

\section{Fast and exact emulation}\label{sec:emulation}

The one-step-ahead emulation framework can be interpreted as a probabilistic analog of the Euler method \citep{biswas2013discussion} for solving ODEs. The classical Euler scheme approximates the evolution of a dynamical system by discretizing time and iteratively applying
\[
\mathbf{x}(t_{s+1}) = \mathbf{x}(t_s) + \Delta t \cdot h(\mathbf{x}(t_s)),
\]
where $h$ denotes the function governing the dynamics of the system. In our setting, this transition is modeled as a data-driven one-step mapping $x(t_s) \to x(t_{s+1})$, learned using GP emulators. Importantly, unlike previously established findings \citep{girard2002gaussian, mohammadi2019emulating} that rely on MC approximation, our method analytically derives the predictive mean and variance using linked GPs, enabling exact and sampling-free uncertainty propagation at each time step. This allows the GP-based flow map to serve as a probabilistic surrogate for the unknown vector field $h$, yielding a data-driven and uncertainty-aware generalization of the Euler scheme. 

\begin{figure}[ht]
    \centering
    \resizebox{0.95\textwidth}{!}{
\begin{tikzpicture}[
    node distance=1cm,
    every node/.style={circle, draw, fill=green!20, minimum size=2.5cm, align=center},
    plain/.style={draw=none, fill=white!20, circle=false, inner sep=0pt},
    arrow/.style={-Stealth, thick}
]
\node (X) [fill=yellow!15] at (0, 0) {$\mathbf{x}(t_0)$};
\node (GP11) [right=of X, yshift=3cm, fill=blue!15] {$\hat{f}_1(\mathbf{x}(t_0))$};
\node [plain, right=of X] (GP12) {$\vdots$}; 
\node (GP13) [right=of X, yshift=-3cm, fill=orange!15] {$\hat{f}_d(\mathbf{x}(t_0))$};
\node (GP21) [right=of GP11, fill=blue!15] {$\hat{f}_1(\mathbf{x}(t_1))$};
\node [plain, right=of GP12] (GP22) {$\vdots$}; 
\node (GP23) [right=of GP13, fill=orange!15] {$\hat{f}_d(\mathbf{x}(t_1))$};
\node [plain, right=of GP21] (GP31) {$\cdots$};  
\node [plain, right=of GP22] (GP32) {$\cdots$};  
\node [plain, right=of GP23] (GP33) {$\vdots$};  
\node (Y1) [right=of GP31, fill=blue!15] {$\hat{f}_1(\mathbf{x}(t_{T-1}))$};
\node (Y2) [plain, right=of GP32] {$\vdots$}; 
\node (Y3) [right=of GP33, fill=orange!15] {$\hat{f}_d(\mathbf{x}(t_{T-1}))$};
\node (xt) [right=of Y2, fill=green!15] {$\mathbf{x}(t_{T})$};
\draw [arrow] (X) -- (GP11);
\draw [arrow] (X) -- (GP12);
\draw [arrow] (X) -- (GP13);
\draw [arrow] (GP11) -- (GP21);
\draw [arrow] (GP11) -- (GP22);
\draw [arrow] (GP11) -- (GP23);
\draw [arrow] (GP12) -- (GP21);
\draw [arrow] (GP12) -- (GP22);
\draw [arrow] (GP12) -- (GP23);
\draw [arrow] (GP13) -- (GP21);
\draw [arrow] (GP13) -- (GP22);
\draw [arrow] (GP13) -- (GP23);
\draw [arrow] (GP21) -- (GP31);
\draw [arrow] (GP21) -- (GP32);
\draw [arrow] (GP21) -- (GP33);
\draw [arrow] (GP22) -- (GP31);
\draw [arrow] (GP22) -- (GP32);
\draw [arrow] (GP22) -- (GP33);
\draw [arrow] (GP23) -- (GP31);
\draw [arrow] (GP23) -- (GP32);
\draw [arrow] (GP23) -- (GP33);
\draw [arrow] (GP31) -- (Y1);
\draw [arrow] (GP32) -- (Y1);
\draw [arrow] (GP33) -- (Y1);
\draw [arrow] (GP31) -- (Y2);
\draw [arrow] (GP32) -- (Y2);
\draw [arrow] (GP33) -- (Y2);
\draw [arrow] (GP31) -- (Y3);
\draw [arrow] (GP32) -- (Y3);
\draw [arrow] (GP33) -- (Y3);
\draw [arrow] (Y1) -- (xt);
\draw [arrow] (Y2) -- (xt);
\draw [arrow] (Y3) -- (xt);
\end{tikzpicture}
    }
    \caption{An illustration of the self-linked GP emulator for a dynamic system governed by $d$ ODEs. Starting with the initial state $\mathbf{x}(t_0)$, each emulator $\hat{f}_m$, $m=1,\ldots,d$ predicts the $m$-th component of the state $x_m(t_{s})$ for $s=1, \ldots, T$. The predicted state $\mathbf{x}(t_{s})$ is then fed forward as the input for the next time step $s+1$.}
    \label{fig:structure}
\end{figure}
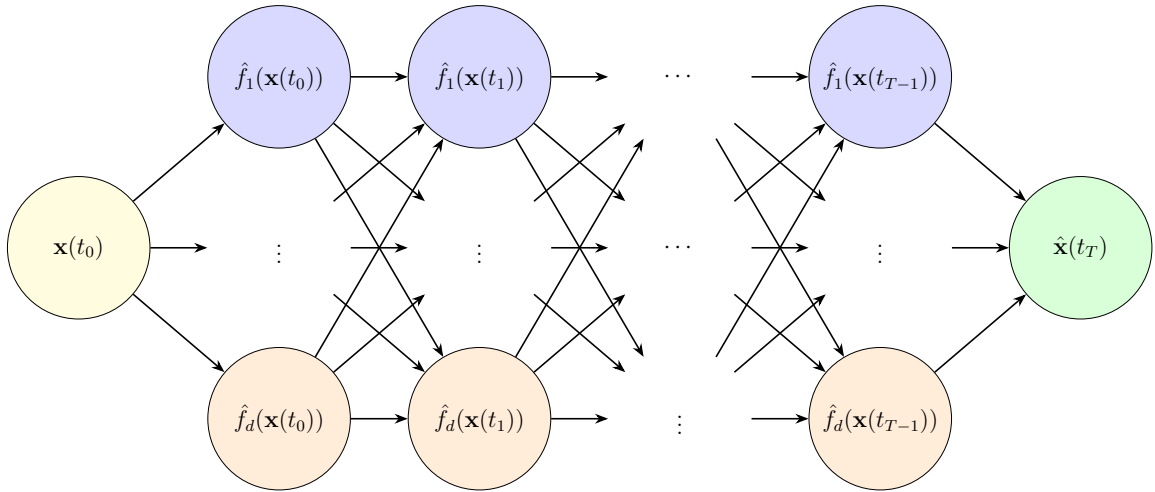

Building on this foundation, we integrate the concept of linked GP emulators to propagate predictions with uncertain inputs over time. Specifically, our method recursively links GP outputs to emulate each state variable, preserving accuracy while maintaining computational efficiency. The feed-forward architecture of the proposed framework, illustrated in Figure \ref{fig:structure}, forms a sequentially layered structure where each predicted state is linked to the previous one. Interestingly, this structure can be interpreted as a special case within the emerging field of deep GP (DGP) models \citep{damianou2013deep}, which have gained significant attention in the computer experiments literature for their hierarchical representation and enhanced flexibility \citep{ko2022deep, sauer2023active, ming2023deep}. However, unlike canonical DGPs that involve latent intermediate layers and require approximate inference, our approach replaces those latent layers with a series of GP emulators, linked recursively to propagate states in a feed-forward manner reminiscent of deep learning models.

We now detail the proposed exact and sampling-free, recursive emulation procedure for the dynamic simulator $f$ introduced in Section \ref{sec:onestep}. As specified in equation \eqref{eq:markov}, the simulator evolves via the one-step transition $\mathbf{x}(t_{s+1})=(\hat{f}_1(\mathbf{x}(t_{s}),\mathbf{w}(t_{s})), \ldots, \hat{f}_d(\mathbf{x}(t_{s}),\mathbf{w}(t_{s})))^\top$, where the forcing input $\mathbf{w}(t_{s})$ may be omitted in autonomous systems. To emulate this mapping, we construct $d$ independent GP models $\hat{f}_m$, one for each state variable $m=1,\ldots,d$, using the one-step-ahead responses $\mathbf{y}_m=(x_m^1(t_1),\ldots,x_m^n(t_1))$. Depending on the presence of forcing inputs, the emulator takes the form either $x_m(t_{s+1})=\hat{f}_m(\mathbf{x}(t_s),\mathbf{w}(t_s))$ or $x_m(t_{s+1})=\hat{f}_m(\mathbf{x}(t_s))$. 

Predictions are obtained recursively using the following closed-form expressions for the posterior predictive mean in \eqref{eq:totalexp} and variance in \eqref{eq:totalvar} under a squared exponential kernel, originally derived by \cite{kyzyurova2018coupling} and adapted here for the one-step-ahead emulation setting:
\begin{align}
&\mathbb{E}(x_m(t_{s+1}))=
\begin{pmatrix}
1 & \mathbf{x}(t_{s})
\end{pmatrix}
\beta_m \nonumber \\
&\quad + \sum^{n}_{i=1} r_i \prod_{j=1}^{p} \exp\left( -\frac{\left(w^i_j(t_0)-w_j(t_{{s+1}})\right)^2}{\theta_{wj}} \right) \prod_{l=1}^d \sqrt{\frac{\theta_{ml}}{\theta_{ml}+2\mathbb{V}(x_l(t_{s}))}} \exp{\left( -\frac{\left(x^i_l(t_0)-\mathbb{E}(x_l(t_{s}))\right)^2}{\theta_{ml}+2\mathbb{V}(x_l(t_{s}))} \right)}, \label{eq:closedmean}
\end{align}
and
\begin{align}
&\mathbb{V}(x_m(t_{s+1})) = \tau^2_m - \left(\mathbb{E}(x_m(t_{s+1}))-
\begin{pmatrix}
1 & \mathbf{x}(t_{s})
\end{pmatrix}
\beta_m \right)^2 \nonumber \\
&\quad + \sum_{i,k=1}^{n} \left(r_i r_k - \tau^2_m (\mathbf{K}^{-1}_m)_{ik} \right) \prod_{j=1}^{p} \exp{ \left(-\frac{\left(w^i_j(t_0)-w_j(t_{s+1})\right)^2+\left(w^k_j(t_0)-w_j(t_{s+1})\right)^2}{\theta_{wj}}\right)} \prod_{l=1}^d\zeta_{lik}, \label{eq:closedvar}
\end{align}
where $r_i = (\mathbf{K}^{-1}_m (\mathbf{y}_m-\begin{pmatrix} 1 & \mathbf{X} \end{pmatrix}\beta_m) )_i$ and
\[
\zeta_{lik} = \sqrt{\frac{\theta_{ml}}{\theta_{ml}+4\mathbb{V}(x_l(t_{s}))}} \exp{\left( -\frac{ \left(\frac{x^i_l(t_0)+x^k_l(t_0)}{2}-\mathbb{E}(x_l(t_{s}))\right)^2}{\frac{\theta_{ml}}{2}+2\mathbb{V}(x_l(t_{s}))} -\frac{(x^i_l(t_0)-x^k_l(t_0))^2}{2\theta_{ml}} \right)}.
\]
Here, $\theta_{ml}$ and $\theta_{wj}$ are the lengthscale hyperparameters associated with the state and forcing input dimensions, respectively.

\begin{algorithm}
\caption{Exact emulation of dynamic simulators}\label{alg:algorithm}
\begin{algorithmic}[1]
\For{$m = 1$ to $d$}
    \State Fit $\hat{f}_m$ on training data $(\mathbf{X}, \mathbf{W}, \mathbf{y}_m)$
\EndFor
\State Set $\mathbf{x}(t_0) \gets \mathbf{x}(t_0)$
\For{$s = 0$ to $T-1$}
    \For{$m = 1$ to $d$}
        \State Compute $\mathbb{E}(x_m(t_{s+1}))$ using \eqref{eq:closedmean}
        \State Compute $\mathbb{V}(x_m(t_{s+1}))$ using \eqref{eq:closedvar}
    \EndFor
    \State Update $s \gets s+1$
\EndFor
\State \textbf{Return:} $\mathbf{x}(t_s)$ for $s=1,\ldots,T$
\end{algorithmic}
\end{algorithm}

The forcing inputs are treated as known and contribute deterministically through their kernel terms, without introducing additional uncertainty. When forcing inputs are absent, the expressions simplify by setting $p = 0$. The closed-form expressions for the posterior predictive mean and variance under a Mat\'ern kernel can be straightforwardly derived following the development in \cite{ming2021linked}. 

Consistent with prior studies \citep{girard2002gaussian, kyzyurova2018coupling, mohammadi2019emulating, ming2021linked}, we adopt the moment matching method to approximate the posterior distribution as Gaussian, leveraging the analytic mean and variance in \eqref{eq:closedmean} and \eqref{eq:closedvar}. This approach is particularly well-suited to the self-coupled structure of the proposed framework, where the prediction at each step serve as input to the next time step. 

By computing the mean and variance analytically, the proposed sampling-free method achieves exact moment propagation, avoiding the accumulation of sampling error. This is especially beneficial for predictions over long time horizon, where even small numerical inaccuracies from early steps can compound over time, leading to significant degradations in later stages—a phenomenon commonly referred to as the \textit{butterfly effect}. Moreover, the feed-forward structure supports efficient multi-step-ahead predictions without repeated recomputation of previous states, providing further computational gains.

\section{Numerical Studies}\label{sec:studies}

In this section, we conduct a series of numerical experiments to evaluate the performance of the proposed approach. Specifically, we compare the predictive performance of the proposed exact algorithm with an emulation method based on MC approximation, considering three sample sizes $(n_{MC}=10, 100, 1000)$. Importantly, both approaches adopt the moment-matching method, and employ the same underlying GP emulator, constructed using an anisotropic squared exponential kernel and a linear mean function $\alpha(\mathbf{x})=(1, \mathbf{x}) \beta$. The difference between the two methods lies solely in how predictive moments are computed as described in Section \ref{sec:review}: the MC-based approach relies on sampling, whereas the proposed exact method computes them analytically.

All numerical experiments are conducted using a maximin Latin hypercube design \citep{johnson1990minimax, morris1995exploratory}, with the number of initial samples set to $n=10 \times d$, following the guidelines of \cite{jones1998efficient} and \cite{loeppky2009choosing}. The time step difference is set to $\Delta t=t_s-t_{s-1}=0.01$.

The governing ODEs of the dynamic simulators are solved using the \textsf{lsoda} method from the \textsf{R} package \textsf{deSolve} \citep{soetaert2010solving}. This solver, known as the Livermore solver for stiff and non-stiff ODE systems, automatically switches between stiff and non-stiff methods for efficiency \citep{petzold1983automatic}.

Predictive performance is evaluated across all time points $t_s$, $s=1,\ldots,T$, using two standard metrics: the root-mean-square error (RMSE) and the mean absolute error (MAE)
, averaged over 100 repetitions. 
The evaluation metrics are defined as follows:
\begin{align*}
    \text{RMSE} &= \sqrt{ \frac{1}{T} \sum_{s=1}^T \left( x_m(t_s) - \mathbb{E}(x_m(t_s)) \right)^2 } ,\\
    \text{MAE} &= \frac{1}{T} \sum_{s=1}^T  | x_m(t_s) - \mathbb{E}(x_m(t_s)) |.
\end{align*}
These metrics measure the discrepancy between the true function and the predictive model, with lower RMSE and MAE 
values indicating higher accuracy. Furthermore, computational efficiency is assessed by comparing computation times across different emulation approaches. 

In addition to prediction accuracy, we assess how predictive uncertainty accumulates over time. Specifically, we determine the number of steps for which predictions remain reliable and the point at which uncertainty increases substantially. For this purpose, we apply change point detection \citep{killick2012optimal} to the predictive standard deviation using the \textsf{cpt.mean} function in the \textsf{changepoint} package \citep{killick2014changepoint}. The comparison is performed only between the exact method and the MC approach with 1000 samples, as the latter provides the most accurate MC-based predictions and yields more stable results by reducing variability from sampling. This analysis addresses the practical question of how far ahead the emulator can provide useful predictions before the uncertainty becomes large enough to degrade performance.

\subsection{Lotka-Volterra model}

The Lotka-Volterra model \citep{lotka1925elements, volterra1927fluctuations, goel1971volterra}, also known as the consumer-prey model, is a foundational framework in mathematical ecology for describing predator-prey dynamics. This model characterizes the temporal evolution of two interacting populations: a prey species with population density $P(t)$ and a predator species with population density $C(t)$. The dynamics are governed by the following system of nonlinear differential equations:
\begin{align*}
\begin{cases}
    \frac{dP}{dt}&=r_G \cdot P \cdot \left( 1-\frac{P}{K} \right) - r_I \cdot P \cdot C,\\
    \frac{dC}{dt}&= k_{AE} \cdot r_I \cdot P \cdot C - r_M \cdot C,
    \end{cases}
\end{align*}
where $r_G>0$ represents the intrinsic growth rate of the prey in the absence of predators, $K$ is the prey carrying capacity, $r_I>0$ is the predation rate coefficient, $k_{AE}>0$ is the assimilation efficiency of the predator, and $r_M>0$ denotes the natural mortality rate of the predator in the absence of prey. For this study, we consider a simulation over $t \in [0,30]$, with prespecified parameter values $r_G = 1.5$, $K = 10$, $r_I = 1.5$, $k_{AE} = 1$, and $r_M = 2$. Initial samples are drawn from $[0,5]^2$. The objective is to predict the system's dynamics given the initial state $\mathbf{x}(t_0) = (P(t_0)=1, C(t_0)=2)$.

\begin{figure}[!t]
\begin{center}
\includegraphics[width=1\textwidth]{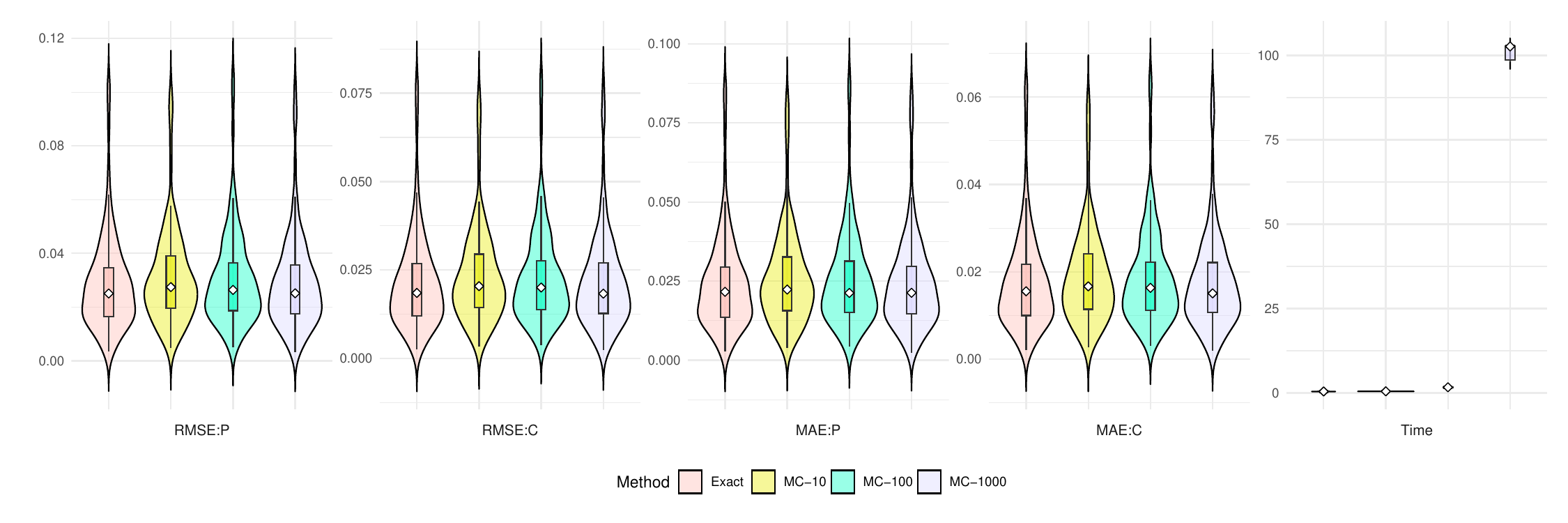} 
\end{center}
\caption{RMSE (first and second panels), MAE (third and fourth panels) of emulation algorithms to $P$ and $C$ and computation time (fifth panel) across 100 repetitions for the Lotka-Volterra model.}
\label{fig:C-P boxplot}
\end{figure}

Figure \ref{fig:C-P boxplot} presents the emulation results over 100 repetitions for RMSE and MAE. As expected, the MC approach produces suboptimal performance, although larger sample sizes yield results closer to those of the exact method. Compared with the exact method, the MC approach with 10 samples yields RMSE and MAE values that are 6–8\% larger, with 100 samples yields values that are 4–6\% larger, and with 1000 samples yields values that are less than 1\% larger. Moreover, their computational costs differ considerably. On average, the exact algorithm requires only 0.45031 seconds per repetition, whereas the MC approach takes 0.48785 seconds with 10 samples, 1.70622 seconds with 100 samples, and 101.37841 seconds with 1000 samples. This comparison highlights the advantage of the proposed framework in significantly reducing computational cost while maintaining accuracy in dynamic system emulation. As a result, the computational gap would increase further if the MC approach employed a larger number of samples, or if the simulation involved many time steps or a longer time horizon.

\begin{figure}[!t]
\begin{center}
\includegraphics[width=1\textwidth]{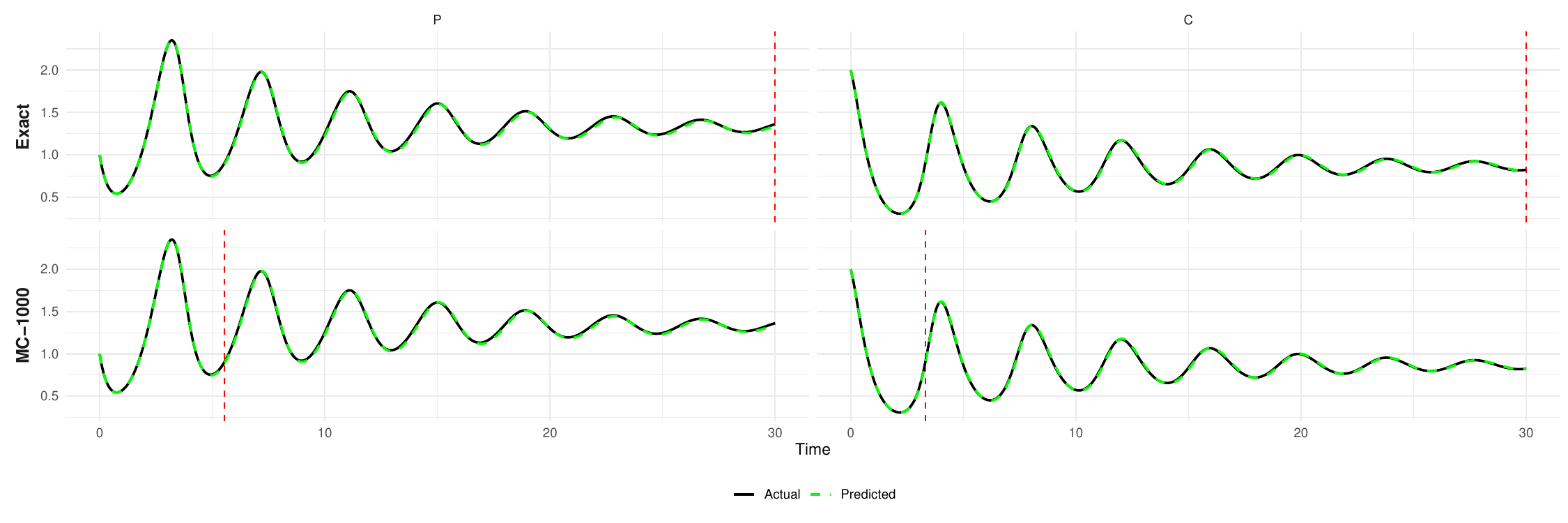} 
\end{center}
\caption{Lotka-Volterra model (black solid line) and emulation results of prediction (green dashed line) via each algorithm with detected change points (red dotted vertical line)}.
\label{fig:C-P changepoint}
\end{figure}

Figure \ref{fig:C-P changepoint} illustrates the predicted trajectories for state variables, $P$ and $C$, along with the change point of predictive uncertainty. Because the one-step-ahead approach is effective enough for emulating the Lotka-Volterra model, both methods accurately predict the simulator outputs with very small uncertainties of approximately $10^{-4}$ for the exact method and $10^{-2}$ for MC method. The Exact method does not detect any change points within the given time steps, whereas the MC method detects changes in uncertainties at $t=5.53$ for $P$ and $t=3.33$ for $C$.

\subsection{Duffing oscillator}
The Duffing oscillator \citep{duffing1918erzwungene, ueda1991survey} is a canonical model in nonlinear dynamics that describes the behavior of a damped oscillator subject to a nonlinear restoring force. It captures the dynamics of a second-order system defined in terms of the displacement variable $x(t)$ and its velocity $v(t)$, governed by the following system of differential equations:
\begin{align*}
\begin{cases}
    \frac{dx}{dt}&=v,\\
    \frac{dv}{dt}&= -\delta v - \alpha x -\beta x^3,
    \end{cases}
\end{align*}
where $\delta > 0$ is the damping coefficient, $\alpha$ controls the linear stiffness, and $\beta$ governs the strength of the nonlinear restoring force. In this study, we simulate the system over $t \in [0,30]$ with parameter values $\delta=0.1$, $\alpha=1$, and $\beta=1$. The initial state is set as $\mathbf{x}(t_0)=(x(t_0)=1, v(t_0)=0)$, and training inputs are sampled from the domain $[-1,1]^2$.

\begin{figure}[!t]
\begin{center}
\includegraphics[width=1\textwidth]{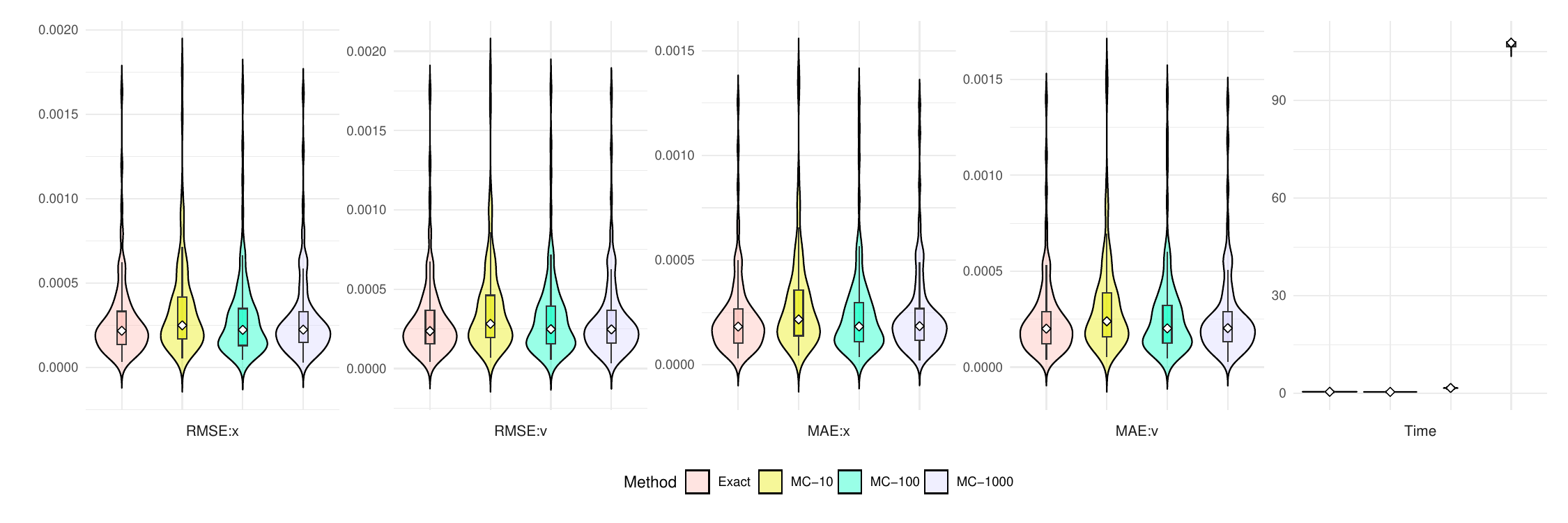} 
\end{center}
\caption{RMSE (first and second panels), MAE (third and fourth panels) of emulation algorithms to $x$ and $v$ and computation time (fifth panel) across 100 repetitions for the Duffing oscillator.}
\label{fig:x-v boxplot}
\end{figure}

Figure \ref{fig:x-v boxplot} illustrates the Duffing oscillator results, which exhibit patterns similar to those observed for the Lotka–Volterra model. The MC approach again underperforms relative to the exact method, although the performance gap narrows as the number of samples increases. With 10 MC samples, RMSE and MAE are approximately 24–25\% larger than those of the exact method; with 100 samples, the difference drops to 5–6\%, and with 1000 samples, to about 2.5\%. Computational times show the same trend as before, with the exact algorithm averaging 0.52205 seconds per repetition compared to 0.47833, 1.69159, and 107.25105 seconds for the MC approach with 10, 100, and 1000 samples, respectively.

\begin{figure}[!t]
\begin{center}
\includegraphics[width=1\textwidth]{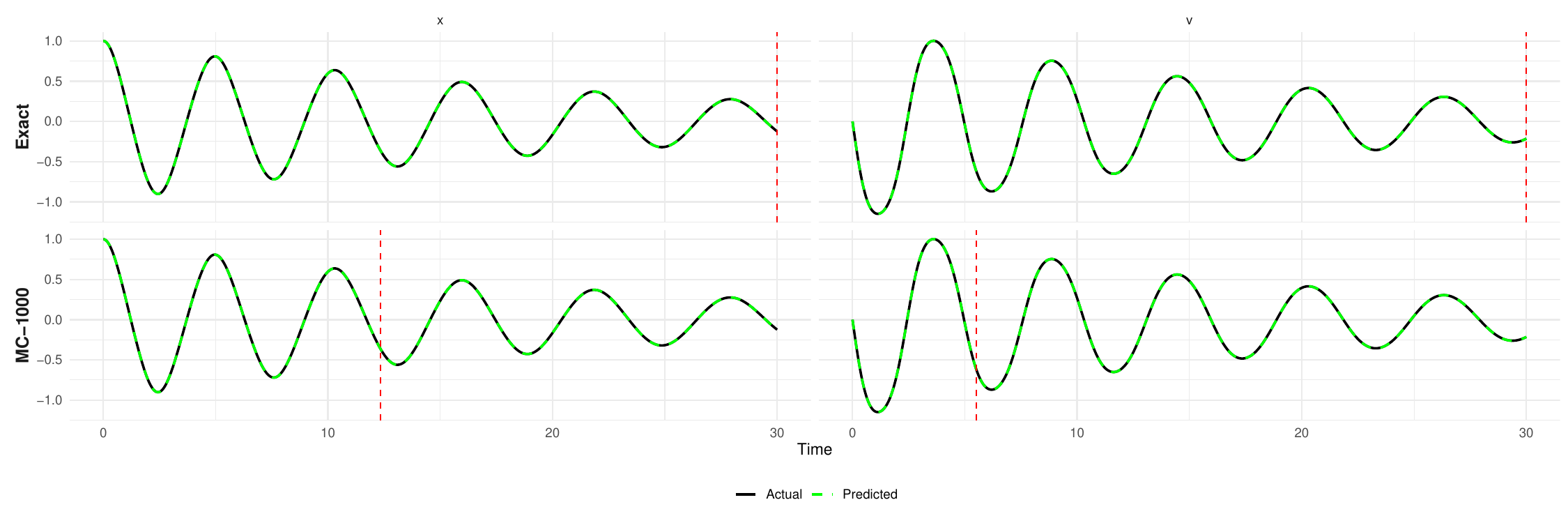} 
\end{center}
\caption{Duffing oscillator (black solid line) and emulation results of prediction (green dashed line) via each algorithm with detected change points (red dotted vertical line)}.
\label{fig:Duffing changepoint}
\end{figure}

Again, the change point detection results for the Duffing oscillator in Figure \ref{fig:Duffing changepoint} follow the same pattern observed for the Lotka–Volterra model. Both methods yield accurate predictions, with predictive uncertainties around $10^{-6}$ for the exact method and $10^{-4}$ for the MC method. As in the previous case, the exact method does not detect any change points, whereas the MC method identifies changes at $t=12.35$ for $x$ and $t=5.51$ for $v$.

\subsection{Lorenz system}

The Lorenz system, introduced by \cite{lorenz1963}, models the behavior of atmospheric convection through a set of three differential equations:
\begin{align*}
\begin{cases}
    \frac{dX}{dt}&=aX+YZ,\\
    \frac{dY}{dt}&=b(Y-Z),\\
    \frac{dZ}{dt}&=-XY+cY-Z,
    \end{cases}
\end{align*}
where $X$, $Y$, and $Z$ represent the system's state variables, and $a$, $b$, and $c$ are system parameters that govern the system’s dynamics. Using the classic values $a=-8/3$, $b=-10$, and $c=28$, originally employed by Lorenz to study convection rolls, the system exhibits chaotic behavior. We focus on a simulation over $t \in [0,25]$, and the initial state $\mathbf{x}(t_0) = (X(t_0)=-1, Y(t_0)=-1, Z(t_0)=-1)$. Initial designs are sampled from $[-10,10]^3$. 

\begin{figure}[!t]
\begin{center}
\includegraphics[width=1\textwidth]{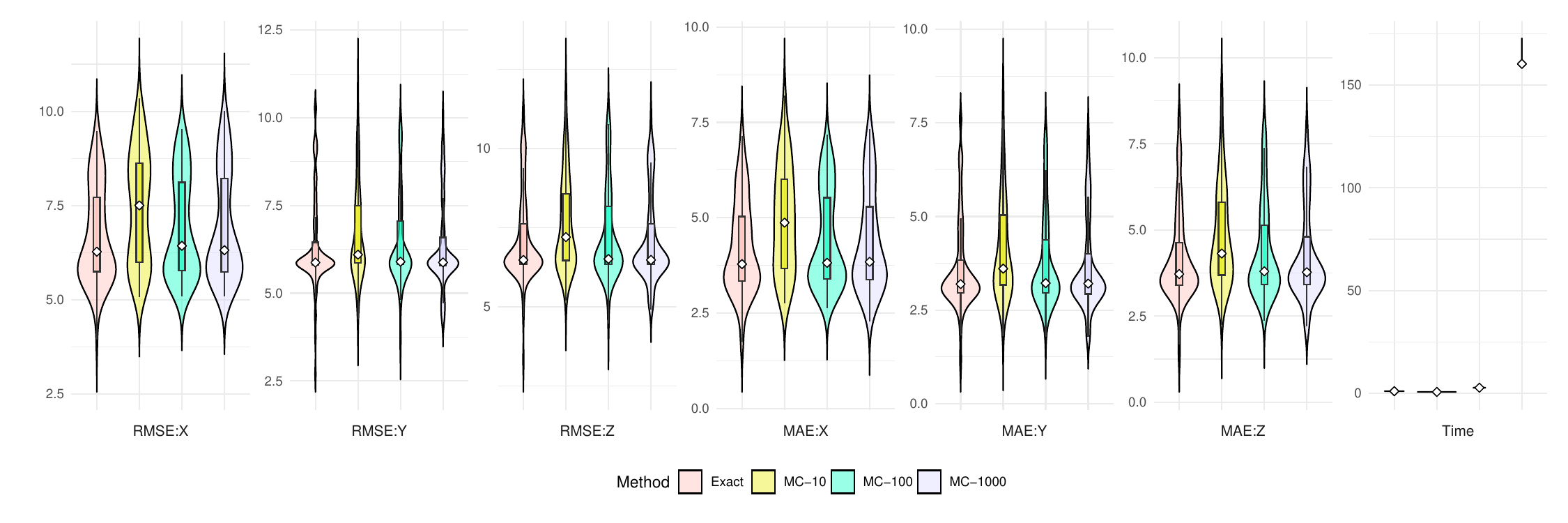} 
\end{center}
\caption{RMSE (first to third panels), MAE (fourth to sixth panels) of emulation algorithms to $X$, $Y$ and $Z$ and computation time (seventh panel) across 100 repetitions for Lorenz system.}
\label{fig:Lorenz boxplot}
\end{figure}

Figure \ref{fig:Lorenz boxplot} shows that the Lorenz system results differ notably from those of the previous two simulators. The Lorenz system is a more sensitive and complex dynamical system, where small changes can lead to large deviations, a phenomenon known as the ``butterfly effect''. This greater sensitivity makes the relative differences between the exact method and MC approaches more pronounced at small sample sizes and more inconsistent across state variables. With 10 MC samples, RMSE and MAE are 6–17\% larger than those of the exact method, with 100 samples, the difference decreases to 1.5–4\%, and with 1000 samples, it ranges from 0\% to 3.5\%. Computational costs remain substantially different, with the exact algorithm requiring 0.99940 seconds per repetition compared to 0.65037, 2.69461, and 160.60290 seconds for the MC approach with 10, 100, and 1000 samples, respectively. These results highlight that the MC approach requires a sufficiently large number of samples to achieve performance comparable to the exact method, but this comes at a substantial increase in computational time.

\begin{figure}[!t]
\begin{center}
\includegraphics[width=1\textwidth]{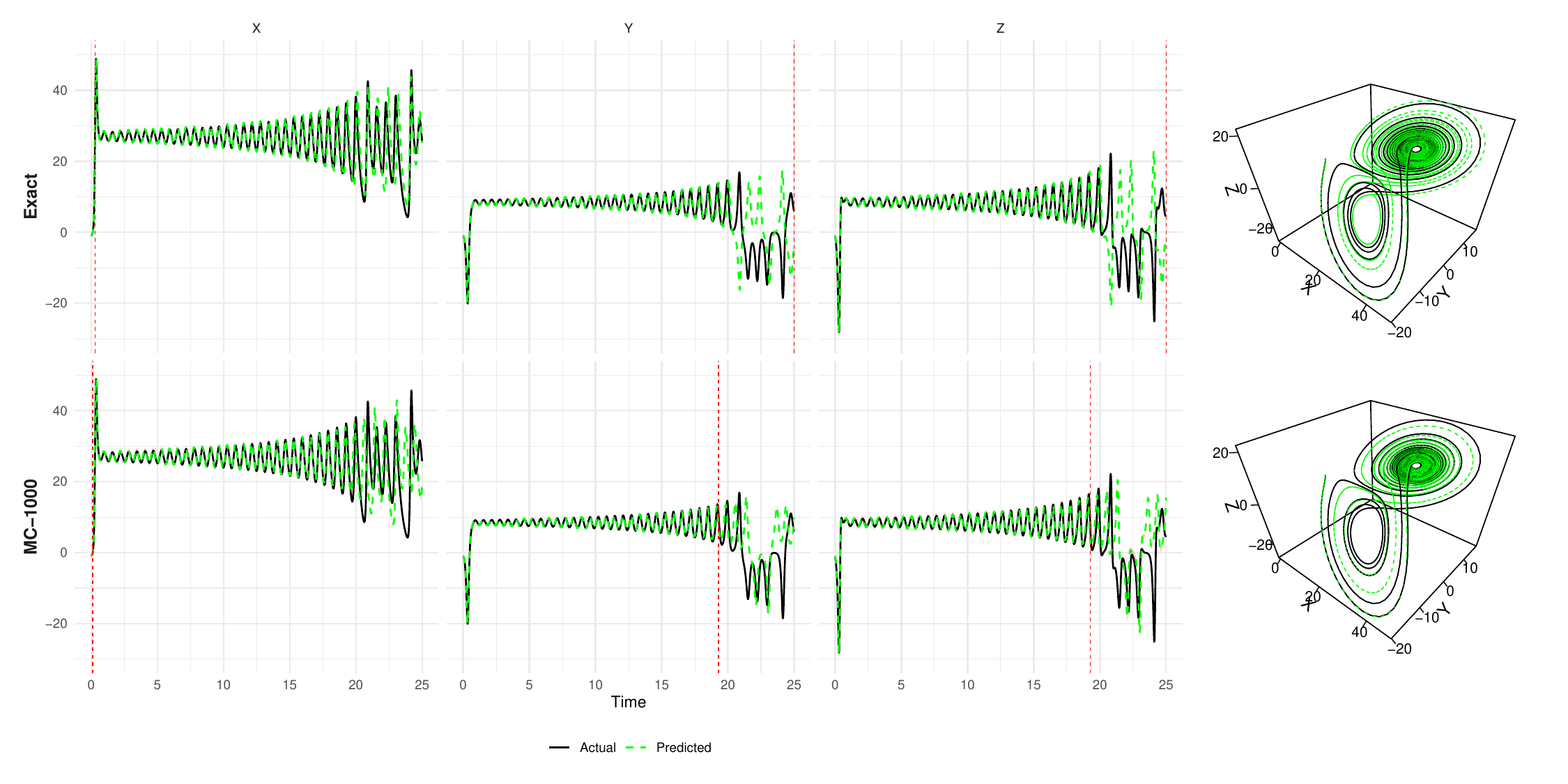} 
\end{center}
\caption{Lorenz (black solid line) and emulation results of prediction (green dashed line) via each algorithm with detected change points (red dotted vertical line) and corresponding 3-D trajectory illustrations.}
\label{fig:Lorenz changepoint}
\end{figure}

Figure \ref{fig:Lorenz changepoint} illustrates the predicted trajectories with change point detection for each state variable, $X$, $Y$, and $Z$. Both algorithms produce accurate predictions at early time steps up to $t=20$, but performance deteriorates thereafter. For the exact method, the first time step at which the absolute difference between the prediction and the true output exceeds 10 is $t=19.88$ for $X$, $t=20.69$ for $Y$, and $t=19.99$ for $Z$, which is later than the corresponding times for the MC method, which are $t=19.77$ for $X$, $t=20.47$ for $Y$, and $t=19.90$ for $Z$. For highly sensitive dynamic systems such as the Lorenz system, the exact algorithm is particularly advantageous, as MC methods introduce variations that exacerbate prediction errors over time. This is consistent with the change point detection results, where the MC approach identifies a change point at $t=0.12$ for $X$ which is near the very beginning of the time steps, and at $t=19.29$ for $Y$ and $t=19.27$ for $Z$. In contrast, exact method maintains smaller uncertainties and detects only a single change point at $t=0.3$ for $X$ caused by increased uncertainty at the peak behavior, while no change points are detected for $Y$ and $Z$ within the given time steps under the same penalty value in the change point detection analysis.

\section{Discussion}\label{sec:conclusion}

In this paper, we propose fast and exact algorithms for emulating complex dynamic simulators. These algorithms offer substantial improvements over traditional MC-based approaches by retaining their key advantages while significantly reducing computational overhead. Specifically, the proposed method bypasses MC approximations by deriving closed-form expressions for the posterior predictive mean and variance. This not only yields more than 150-fold improvements in computational efficiency without compromising accuracy, but also prevents the accumulation of small differences from sampling errors, thereby ensuring stable propagation over time. The framework also naturally accommodates forcing inputs by treating them as additional input variables within the GP formulation, enabling emulation of a broader class of dynamic systems. While uncertainty quantification is also supported in MC-based methods, our method preserves this capability and provides predictive variance analytically at each time step, supporting reliable uncertainty-aware emulation. To facilitate adoption and reproducibility, we provide an \textsf{R} package, \textsf{dynemu}, available on \textsf{CRAN}, which implements the proposed methodology in a user-friendly and computationally efficient framework.

From a computational perspective, the addition of jitter for numerical stability \citep{ranjan2011computationally}--commonly used to avoid ill-conditioned covariance matrices--may introduce minor discrepancies that could accumulate over time. In dynamic systems that are highly sensitive to small changes, such as those exhibiting the butterfly effect, these numerical artifacts can significantly impact the emulation outcomes. Furthermore, computational efficiency can be further improved by implementing low-level programming languages for operations.

In this work, although the state variables are correlated, we followed the approach of \cite{damianou2013deep} and independently constructed the emulators. However, as suggested by \cite{mohammadi2019emulating}, neglecting these correlations might lead to information loss, potentially affecting predictive accuracy. Future work could address this limitation by incorporating correlations among state variables, further enhancing emulation performance.

Another promising direction for future research lies in the emulation of non-periodic chaotic systems, such as the three-body problem \citep{newton1687philosophiae, musielak2014three}. Chaotic systems are inherently unpredictable and highly sensitive to initial conditions, making them particularly challenging to emulate. To enhance model accuracy in such cases, integrating active learning, where new training points are adaptively selected to improve emulation performance could be a valuable approach for dynamic system emulation.




\bibliography{bib}

\end{document}